# SAXS Structural Characterization of Nanoheterogeneous Conducting Thin Films. A Brief Review of SAXS Theories.

M. Cattani* and M. C. Salvadori
*Institute of Physics, University of São Paulo, C.P. 66318, CEP 05315-970
São Paulo, S.P. Brazil    * Electronic mail:  mcattani@if.usp.br*

F. S. Teixeira
*Polytechnic School, University of São Paulo*
*Avenida Professor Luciano Gualberto, Travessa R-158, CEP 05508-900*
*São Paulo, Brazil*

**Abstract.** The SAXS (small angle X-ray scattering) technique has been used to determine the fundamental structure of conducting thin films fabricated by implanting gold ions into PMMA (polymethylmethacrylate) polymer. We present a brief review of the SAXS theories necessary to determine the structural properties of our nanoheterogeneous films.

## 1. Introduction

Conducting thin films composed by metallic aggregates embedded in a polymeric matrix have been largely studied because of their interesting physical and chemical properties with potential technological applications in the fields of plastic electronics [1], photonics [2,3], and chemical/biosensing [4-6] systems. These composites have been formed using a broad range of techniques, including sputtering, chemical reduction, photo reduction, emulsion, laser vaporization [2-6] and metal ion implantation[3,7-11] into polymer targets. Whatever is the fabrication technique used, the electrical conductivity of such films is very dependent on the relative concentration of metal-polymer and on the way it is arranged. The general rule is that metal nanoparticles form clusters and, as the amount of metal is increased, these aggregates start to connect. When it occurs, an abrupt increase of electrical conductivity happens. This critical point, called percolation threshold [9] occurs when the metal concentration deposition attaints a critical dose $\phi_c$.

In preceding works [9-11] we have fabricated PMMA-Au composite thin films by implanting very low energy (49 eV) Au ions into polymer. We analyzed the composite percolation critical properties with electrical measurements [9,10], the lithographic properties [10] and the surface plasmon resonance effect, as well the size distribution of gold nanoparticles by Transmission Electron Microscopy (TEM) images [11]. The TEM images have shown [11] that the Au-polymers composites are thin films (with an average thickness ~ 7 nm) formed by gold nanoparticles clusters with sizes between 2 nm and 12 nm. According to these images the gold nanoclusters are almost spherical. The clusters formed by very small gold nanoparticles that are usually named *monomers*. However, TEM analysis are not sufficiently accurate to give us detailed information about the sizes of the gold monomers and of the clusters, the geometrical distribution of these monomers inside the clusters and how the clusters are arranged in the aggregate. In order to obtain detailed information about these structural properties it would be necessary to submit our films to the small angle X-ray scattering (SAXS) [12,13]. The X-ray scattering intensity I(**q**) is experimentally determined as a function of the scattering vector **q** whose modulus is given by $q = (4\pi/\lambda)\sin(\theta/2)$, where $\lambda$ is the X-ray wavelength and $\theta$ is the scattering angle. Since our composite films are macroscopically isotropic the intensities will depend only on the modulus of $q$ that for SAXS is given by $q \approx (2\pi/\lambda)\ \theta$. The SAXS technique is useful if relevant structural features are at a superatomic level, from a few tenths up to 100 nm.

As well known, the SAXS measurements analyzed with suitable theoretical models that accounts for disordered systems peculiarities are of fundamental importance in the characterization of nanostructured materials [14]. Aggregates of metallic nanoparticles [15,16], silica [17,18] and globular micelles [19] have been recently analyzed using the SAXS technique. In next sections we present a brief review of the SAXS theories that will be used to analyze our experimental SAXS data in future works.

**(2) Brief Review of SAXS Theories.**

The theoretical approaches will be presented assuming that our films are isotropic and that elementary gold nanoparticles (*monomers*) constituents of the films are identical, spherical with radius $r_o$ and with an average electronic density $\rho$. The films are composed by an ensemble clusters of monomers which generate the total scattered intensity I(q). However, instead of calculating directly the total I(q) we will show how to calculate first the scattering due to individual monomers, after the scattering due to individual clusters and, finally, the total scattering due to the ensemble of all clusters defined as aggregate or sample.

## (2.1) The scattered intensity $I_1(q)$ due to a single monomer.

According, for instance, to Debye and Bueche [20] and Guinier [12] the intensity $I_1(q)$ of the scattered X-rays due to a monomer is given by

$$I_1(q) = I_e(q) \int_v d\mathbf{r} \int_V d\mathbf{r}' \rho(\mathbf{r}) \rho(\mathbf{r}') \exp[i\mathbf{q}\cdot(\mathbf{r} - \mathbf{r}')] \quad (2.1),$$

where $I_e(q)$ is the intensity scattered by one electron, v and $\rho(\mathbf{r})$ are, respectively, the volume and the electronic density of the monomer.

Assuming that $\rho(\mathbf{r})$ has a radial symmetry, that is, $\rho(\mathbf{r}) = \rho(r)$ and taking $\mathbf{q} = q\,\mathbf{z}$ where $\mathbf{z}$ is unit vector along the z-axis, the volume integrals of (2.1) become, using the polar spherical coordinates,

$$\int_v \rho(r)\, r^2 dr\, \sin\theta\, d\theta\, d\varphi\, \exp(iqr\cos\theta) \int_v \rho(r')\, r'^2 dr'\, \sin\theta'\, d\theta'\, d\varphi'\, \exp(iqr'\cos\theta'),$$

that integrating on the variables $\theta$, $\varphi$, $\theta'$ and $\varphi'$ become

$$\int_0^\infty \rho(r)\, 4\pi r^2 dr\, [\sin(qr)/qr] \int_0^\infty \rho(r')\, 4\pi r'^2 dr'\, [\sin(qr')/qr'] .$$

Consequently,

$$I_1(q) = I_e(q) \left\{ \int_0^\infty \rho(r)\, 4\pi r^2 dr\, [\sin(qr)/qr] \right\}^2 \quad (2.2).$$

Assuming that the monomers are spheres with radius $r_o$ and with electronic density $\rho(r)$ = constant = $\rho$ for $r \leq r_o$ and $\rho(r) = 0$ for $r > r_o$ one can easily verify, using (2.2), that

$$I_1(q) = I_e(q)\, F(q)^2 = I_e(q)\{3\rho v_o\, [\sin(qr_o) - qr_o \cos(qr_o)]/(qr_o)^3\}^2 \quad (2.3),$$

where $v_o$ is the monomer volume and the function $F(q)$ is given by

$$F(q) = 3\rho v_o\, [\sin(qr_o) - qr_o \cos(qr_o)]/(qr_o)^3 \quad (2.4),$$

which is named *single-particle form factor*.

A different approach developed originally by Rayleigh [21,12,13] to calculate $I_1(q)$ (using a correlation function for the internal structure of the monomers) is shown in the Appendix.

**(2.2) Monodispersive dilute cluster.**

In the general case, if all scattering particles of the system are identical (same forms, sizes and compositions) we say that it is *monodispersive.* If the scatters of the system have different forms, sizes and compositions it is called *polydispersive.*

Here we will assume that our system is a dilute monodispersive cluster with arbitrary volume V and number $N_o$ of monomers. Assuming that the monomers are separated from each other widely enough it is plausible that they will make independent contributions to the scattered intensity. In these conditions the total scattered intensity due to the cluster will be given by, omitting for simplicity the factor $I_e(q)$,

$$I_o(q) = N_o I_1(q) = N_o \rho^2 v_o^2 P(q) \qquad (2.5),$$

where P(q) is defined by

$$P(q) = \{ 3 [\sin(qr_o) - qr_o \cos(qr_o)]/(qr_o)^3 \}^2$$

Note that (2.5) is valid when the individual scattering objects (embedded in the vacuum) are sufficiently apart so that they can be taken as isolated.

Let us now suppose that the $N_o$ spherical monomers are embedded in a homogeneous medium with constant electric density $\rho_o$. As is known [12,13], a homogeneous medium does not scatter the X-ray; only the difference $\Delta\rho = \rho - \rho_o$ will be relevant for diffraction. In these conditions the total intensity (2.5) will be replaced by

$$I_o(q) = N_o \Delta\rho^2 v_o^2 P(q) \qquad (2.6).$$

**(2.3) Monodispersive dense cluster.**

Equation (2.6) gives the scattering intensity produced by a single cluster composed by $N_o$ monomers widely separated. However [12], when these monomers form a dense cluster (as occurs, for instance, in liquids) (2.6) does not give a satisfactory description of scattering intensity. So, according to the approach proposed by Zernicke and Prins [22] the intensity I(q) due to a dense isotropic cluster is given by [12,17-19]:

$$I(q) = I_o(q) S(q) = I_o(q) \{ 1 + 4\pi\Phi \int_0^\infty [g(r) - 1] r^2 (\sin(qr)/qr) \, dr \} \qquad (2.7),$$

where $I_o(q)$ is given by (2.6) and the function $S(q)$ is named *structure factor of the particles centres*. $g(r)$ is the *pair correlation function* [23,24] defined by

$$g(\mathbf{r}) = \int_V \rho(\mathbf{r})\, \rho(\mathbf{r}+\mathbf{r}')\, d\mathbf{r}' \qquad (2.8),$$

where V is cluster volume and $\Phi = (N_o/V)$. As well known [23,24] $\Phi g(r)$ represents the probability per unit of volume to find a particle at a distance r from a particle situated at the origin. For a non-fractal Euclidean 3-dim aggregate the total number of particles $N(r)$ within a sphere of radius r centered in a central particle is given by

$$N(r) = \Phi \int_0^r g(r)\, 4\pi\, r^2\, dr \qquad (2.9).$$

The function $g(r)$ can be calculated [23,24] taking into account the pair interaction potential $u(r)$ between the particles of the system. Using (2.7) and the predicted $g(r)$ we calculate the structure factor $S(q)$ that can be compared with the same factor obtained from the experimental spectrum $I(q)$ [23,24].
    In this way for a dense isotropic cluster with volume V, composed by $N_o$ spherical particles with radius $r_o$ and constant (or average) electronic density $\rho$ the scattered intensity $I(q)$ is given by

$$I(q) = N_o (\Delta\rho)^2\, v_o^2\, P(q)\, S(q) \qquad (2.10),$$

where $P(q)$ and $S(q)$ are shown by (2.5) and (2.7), respectively.
    Since $N_o(\Delta\rho)^2\, v_o^2 \sim M^2$, where M is the mass of the cluster, (2.10) can be written for dense aggregates as $I(q) \sim M^2\, S(q)$ as will be seen in Section 3.2.

**(2.4) Monodispersive fractal aggregates: dilute and dense clusters.**
    As well known, it is widely recognized [14,25,26] that the complex microstructure and behavior of a large variety of materials and systems can be quantitatively characterized by using the ideas of fractal distributions. Fractal concepts give us an important tool for characterizing the geometry and surface structure of heterogeneous materials and long-range correlations that often exist in their morphology. Even if the material does not possess fractal properties at significant length scales the concepts of fractal geometry often provide useful means of obtaining deeper insights into the structure of the material [14]. So, let us analyze the case of fractal aggregates.

The intensity I(q) depends on two different kinds of fractalities: (A) *boundary surface fractality* of the monomers and of the clusters and (B) *mass fractality* of the monomers ensemble.

(A) The scattering effects due to the *surface fractality* of the monomers in *dilute* clusters have been estimated by Bale and Schmidt [27]. Taking into account the monomer fractal boundary surface they have calculated the scattered intensity due to a spherical monomer. They have shown that the function $I_o(q)$ given by (2.6) must be replaced by a new function $I_o^s(q)$. For large q values the intensity $I_o^s(q)$ is given by:

$$I_o^s(q) \approx \pi N_o (\Delta\rho)^2 \Gamma(5-d) \sin[\pi(D_s-1)/2]/q^{(6-D_s)} \qquad (2.11),$$

where $\Gamma(x)$ is the gamma function of argument x, d is the Euclidian dimension of sample and $D_s$ is the fractal dimension of the monomer surface. In this paper it will be assumed that the monomers have a homogeneous internal structure, with no volume fractality.

(B) Let us see how to evaluate the effects the *mass fractality* of the cluster. From (2.9) we get

$$dN(r) = \Phi g(r) 4\pi r^2 dr \qquad (2.12)$$

A fractal aggregate, within a mathematical context, is characterized by a spatial distribution of the individual scatters given by [26]:

$$N(r) = (r/r_o)^D \qquad (2.13),$$

where D is the mass fractal dimension. Differentiation of (2.13) and identification with (2.12) gives the probability per unit of volume,

$$\Phi g(r) = (D/4\pi r_o^D) r^{D-3} \qquad (2.14).$$

Since D is smaller than 3 [25,26], g(r) given by (2.14) goes to zero at large r, which is clearly unphysical. For large r the sample will show a macroscopic density, uniform, with negligible fluctuations. As is known, according to the theory of liquids [23,24], $g(r) \to 1$ when $r \to \infty$. To eliminate these problems and to avoid the divergence in the evaluation of S(q) it was proposed [18,19,28] to introduce a cut-off distance $\xi$, *named correlation length*, to describe the physical behavior of the pair correlation function at large distances. With these assumptions it was shown that

$$\Phi[g(r) - 1] = (D/4\pi r_o^D) \, r^{D-3} \exp(-\xi/r) \qquad (2.15).$$

The meaning of $\xi$ is only qualitative and has to be made precise in any particular situation. In a general way it represents the characteristic distance above which the mass distribution in the cluster is no longer described by a fractal law. It plays a role quite similar to the *correlation length* associated with the roughness of disordered materials [29,30].

Substituting (2.15) into (2.7) we get the function $S^v(q)$ [18,19]

$$S^v(q) = 1 + (D/r_o^D) \int_0^\infty r^{D-1} \exp(-r/\xi) [\sin(qr)/qr] \, dr$$

$$= 1 + (1/qr_o)^D \{D \, \Gamma(D-1)/[1 + 1/(q\xi)^2]^{(D-1)/2}\} \sin[(D-1)\tan^{-1}(q\xi)] \qquad (2.16).$$

In this way, due to the mass fractality and the surface fractality of the monomers we have now, instead of (2.10):

$$I(q) = I_o^s(q) \, S^v(q) \qquad (2.17),$$

where $S^v(q)$ is given by (2.16) and $I_o^s(q)$ is given elsewhere [27]. For large q values $I_o^s(q)$ is given by (2.11).

According to the momentum uncertainty relation (MUR) $\Delta p \, \Delta r \geq \hbar$ where $\Delta p = \hbar q$ we see that the radius $R \sim \Delta r$ of the region inside which the x-ray scattering is produced can be estimated by $R \sim 1/q$. Let us obtain the function I(q) for some special limits of q.

(A) $q\xi \ll 1$ and, consequently, $qr_o \ll 1$

In this case $I_o(q) \rightarrow N_o(\Delta\rho)^2 v_o^2$ and $S^v(q) \rightarrow \Gamma(D+1)(\xi/r_o)^D\{1 - [D(D+1)/6]q^2\xi^2\}$ resulting a I(q) function which has a Guinier-type behavior

$$I(q) = N_o (\Delta\rho)^2 v_o^2 \, \Gamma(D+1)(\xi/r_o)^D\{1 - [D(D+1)/6]q^2\xi^2\}$$

$$\approx N_o (\Delta\rho)^2 v_o^2 \Gamma(D+1)(\xi/r_o)^D \exp[-D(D+1)\xi^2 q^2/6] \qquad (2.18).$$

According to (2.18) the *generalized gyration radius* $R_g(D,\xi)$ for spherical particles is given by

$$R_g(D,\xi) = [D(D+1)/2]^{1/2} \, \xi \qquad (2.19).$$

instead of $R_g = (3/5)^{1/2} r_o$ predicted by Guinier[12] that is valid when the fractality is negligible. The cluster radius is defined by (2.19). Inside the sphere with radius $R_g$ ~ $\xi$, where the monomers are assembled, there is a correlation effect between them. The correlation length $\xi$ represents the characteristic distance above which the mass distribution is no longer described by a fractal law.

According to the MUR, the momentum transfer q in the cluster region is given approximately by q ~ $1/R_g$ ~ $1/\xi$.

(B) $1/\xi \ll q \ll 1/r_o$

In this q range $I_o(q) \to N_o(\Delta\rho)^2 v_o^2$, $S^v(q) \sim q^{-D}$ and, consequently, $I(q) \sim q^{-D}$. In this region the log I(q) x q plot is generally used to estimate the fractal dimension[17-19] measuring the slope of the curve. For homogeneous materials D = 3 we obtain $I(q) \sim q^{-3}$. Since $D \leq 3$ we verify that as q increases I(q) becomes larger for mass fractal than for homogeneous aggregates.

In this fractal range, that is, when $1/\xi \ll q \ll 1/r_o$, the radius $R$ of the scattering regions are in the interval $\xi > R > r_o$.

(C) $qr_o \gg 1$

In this case $S^v(q) \to 1$ and there are two possibilities for $I_o(q)$: when the surfaces of the monomers are smooth, that is, when $D_s = 2$ and when the surfaces are fractal, that is, when $D_s > 2$.

In the first case, that is, when $D_s = 2$ we verify that $F^2(q) \to 9(qr_o)^{-4}/4$. This implies that I(q) becomes,

$$I(q) = 2\pi(\Delta\rho)^2 S_o/q^4 \qquad (2.20),$$

where $S_o = N_o \pi r_o^2$ is the total area of the monomers. Equation (2.20) which predicts a intensity $I(q) \sim 1/q^4$ is known as *Porod law* [12,13].

In the second case, that is, when $D_s > 2$ we get, using (2.11):

$$I(q) \approx \pi N_o (\Delta\rho)^2 \Gamma(5-d) \sin[\pi(D_s - 1)/2]/q^{(6-Ds)} \qquad (2.21),$$

which is the *generalized Porod law* showing that, for large q values, due to the surface monomers fractality the intensity $I(q) \sim 1/q^{(6-Ds)}$, instead of $I(q) \sim 1/q^4$. Note that for $D_s = 2$ (smooth surfaces) (2.21) becomes equal to (2.20). As $3 > D_s \geq 2$, the q exponent $6-D_s$ falls between 3 and 4 [14,18,19] the $1/q^4$ dependence is not easily distinguishable from the smooth surface with $D_s = 2$. Anyway, the surface fractality tends to increase I(q) as q increases.

## (3) Monodispersive and polydispersive fractal aggregates

In Section 2.4 we have shown how to calculate the intensity I(q) due to a dense fractal isotropic cluster formed by $N_o$ identical spherical monomers with radius $r_o$ and uniform electronic density $\rho$. The cluster radius is defined the *radius of gyration* $R_g = [D(D+1)/2]^{1/2}\,\xi$.

We analyze now I(q) due to an *ensemble* (*aggregate*) of clusters. We say that an aggregate is *monodispersive* is when it composed by N identical clusters: have the same radius and the same number $N_o$ of identical monomers. In this paper we say that an aggregate is *polydispersive* when it is composed by N clusters that can have different $R_g$ and different number of identical monomers. All clusters are assumed to have the same mass fractal dimension D. The monomers surfaces have fractal dimension $D_s$.

(3.1) Polydispersive dilute aggregate.

Thus, let us assume that our polydispersive aggregate is formed by a **dilute** collection of N fractal clusters which have radius $\{R_i\}_{i=1,...,N}$, each one with $N_{oi}$ monomers with **smooth** surfaces ($D_s = 2$). Indicating by $N(i)_{i=1,...,N}$ the number of clusters with radius $R_i$ we see, following the formalism presented in Section 2, that the SAXS intensity generated by this polydispersive aggregate is proportional to $I_{pol}(q)$ that is given by:

$$I_{pol}(q) = \sum_{i=1..N} N(i)\, I_i(q) \qquad (3.1),$$

where, according to (2.6) and (2.16), $I_i(q) = I_{oi}(q)\, S_i^v(q)$,

$$I_{oi}(q) = N_{oi}\,(\Delta\rho)^2\, v_o^2\, P(q) = N_{oi}\,(\Delta\rho)^2\, v_o^2\, \{3\,[\sin(qr_o) - qr_o \cos(qr_o)]/(qr_o)^3\}^2$$

and $\qquad\qquad(3.2)$

$$S_i^v(q) == 1 + (1/qr_o)^D \{D\,\Gamma(D-1)/[\,1 + 1/(q\xi_i)^2]^{(D-1)/2}\}\,\sin[(D-1)\tan^{-1}(q\,\xi)]\,.$$

Note that all clusters have the same fractal dimension D but different clusters have a different correlation length $\xi_i$ since $R_i = [D(D+1)/2]^{1/2}\,\xi_i$.

Taking into account that the factor $(\Delta\rho)^2\, v_o^2\, P(q)$ is the same for all clusters $I_{pol}(q)$, given by (3.1), can be put in a more compact form:

$$I_{pol}(q) = (\Delta\rho)^2\, v_o^2\, P(q) \sum_{i=1..N} N(i)\, N_{oi}\, S_i^v(q) \qquad (3.3).$$

For a monodispersive sample formed by N identical clusters the intensity $I_{mon}(q)$, using (4.3) is simply given by

$$I_{mon}(q) = N\, N_o\, (\Delta\rho)^2\, v_o^2\, P(q)\, S^v(q) \qquad (3.4).$$

Now, let us apply (3.3) to study the simplest polydispersive aggregate formed only by two clusters with radius $R_1$ and $R_2$. Supposing $R_1 > R_2$ we have $\xi_1 > \xi_2$. In the q region where $q\xi_1 \ll 1$ and $qr_o \ll 1$ we have

$$I_1(q) \sim \Gamma(D+1)(\xi_1/r_o)^D \exp[-R_1^2 q^2/3]\,, \qquad (3.5)$$

where $R_1 = [D(D+1)/2]^{1/2}\xi_1$, according to (2.18) and (2.19), respectively. Similarly, in the region where $q\xi_2 < 1$ and $qr_o \ll 1$ we have

$$I_2(q) \sim \Gamma(D+1)(\xi_2/r_o)^D \exp[-R_2^2 q^2/3]\,, \qquad (3.6)$$

with $R_2 = [D(D+1)/2]^{1/2}\xi_2$. Consequently,

$$I_{pol}(q) \sim f_1 I_1(q) + f_2 I_2(q), \qquad (3.7)$$

where $f_1$ and $f_2$, with $f_1 + f_2 = 1$, are the percentages of clusters 1 and 2, respectively.

With a log $I_{pol}(q)$ x $q^2$ plot [16] we can estimate the clusters radius $R_1$ and $R_2$ measuring the slopes of the curves $I_1(q)$ and $I_2(q)$.

If $f_1$ and $f_2$ are known the equation $I_{pol}(q) \sim P(q)\,[f_1 S_1(q) + f_2 S_2(q)]$ can be used to fit the experimental data $I(q)$ in terms of the adjustable parameters D, $\xi_1$, $\xi_2$ and $r_o$ that, in this way, can be determined.

(3.2) Polydispersive dense aggregate.

In the general case, the scattered intensity $I_M(q)$ due to a dense cluster with mass M is given by [12,20, 31-34]:

$$I_M(q) \sim M^2 \int_V d\mathbf{r} \int_V d\mathbf{r}'\, \rho(\mathbf{r})\, \rho(\mathbf{r}')\, \exp[i\mathbf{q}\cdot(\mathbf{r}-\mathbf{r}')] = M^2 \int_V d\mathbf{r}\, g(\mathbf{r})\, \exp[i\mathbf{q}\cdot\mathbf{r}] \qquad (3.8),$$

where V and M are the volume and mass of the cluster, $\rho(\mathbf{r})$ is the electronic density of the cluster medium and $g(\mathbf{r})$ is the pair correlation function defined by (2.8). The mass M and radius of gyration R of the cluster are given by,

$$M = \int_V d\mathbf{r}\, g(\mathbf{r}) \qquad \text{and} \qquad R^2 = (1/2M) \int_V d\mathbf{r}\, \mathbf{r}^2\, g(\mathbf{r}) \qquad (3.9).$$

The *structure factor* $S_M(q)$ associated with the cluster with mass M is given by [12,18],

$$S_M(q) = (1/M) \int_V d\mathbf{r}\, g(\mathbf{r}) \exp[i\mathbf{q}\cdot\mathbf{r}] \qquad (3.10).$$

It is important to note that in this dense case does not explicitly appear the scattering due to individual monomers as occurs for dilute aggregates. As seen from (3.8) and (3.10) $I_M(q)$ is proportional only to the *structure factor* $S_M(q)$.

Let us suppose in what follows that the cluster has a spherical symmetry. As was seen in Section (2), the function g(r) for an infinite (mathematical) fractal may be represented by simple power law, a finite (physical) fractal has a natural length scale $\xi$ which can be introduced into g(r) through a crossover function $f(x) = f(r/\xi)$

$$g(r) = f(r/\xi)\, r^{D-3} \qquad (3.11).$$

For $x \ll 1$, $f(x) \sim 1$, whereas for $x \gg 1$, f(x) must decreases faster than a power law like, for instance, $f(x) \sim \exp(-x)$. It may be readily show, using (3.8)-(3.11) that $M = f_{D-3}\,\xi^D$ and $R^2 = \xi^2 f_{2+D-3}/2\,f_{D-3}$ where $f_\alpha = \int f(s)\,s^\alpha\,d^3s$, which gives the necessary fractal relation $M \sim R^D$.

From (3.10) it can be verified [32] that $S_M(q)$ is a function of the variable $q\xi$ alone, that $S_M(q)$ is normalized in the usual way ($S_M(0) = 1$) and that the general small qR expansion, $S_M(q) = 1 - (qR)^2/3 + \ldots$, can be obtained from (3.10) using (3.9). After a long calculation it can be shown [31] that for $qR \gg 1$ the structure factor $S_M(q)$ is given by $S_M(q) \sim (qR)^{-\gamma}$, where $\gamma = -D$. Note that this result is similar to that found in the paragraph (B) of the Section 3 when $qr_o \gg 1$.

In a few words, it was shown [31,32] that for dense mass fractal cluster $I_M(q)$ is given by,

$$I_M(q) \sim M^2 F(qR) \qquad (3.12),$$

where $F(qR) = 1 - (qR)^2/3 + \ldots$ for $qR \ll 1$ and $F(qr) \sim (qR)^{-D}$ for $qR \gg 1$.

Indeed, for $qR \ll 1$ from (4.12) we get, putting $M \sim R^D$.

$$I_M(q) \sim R^{2D}\,[1 - (qR)^2/3], \qquad (3.13)$$

which is similar to the scattered intensity I(q) given by (3.20) and (3.21) taking into account that $M \sim N_o(\Delta\rho)^2 v_o^2$ and $\xi \sim R$.

In the opposite limit, that is, for $qR \gg 1$ we obtain, putting $M \sim R^D$

$$I_M(q) \sim R^{2D}\,(qR)^{-\gamma} = R^D q^{-D} \qquad (3.14),$$

which describes the intensity behavior $I(q) \sim q^{-D}$ in the fractal region predicted by in Section 2.4.

To compute the effect of the polydispersivity, the single cluster intensity $I_M(q)$ must be averaged over the number distribution of clusters $N(M)$ that are present in the aggregate. So, the $I(q)$ due to all clusters of the aggregate is given by,

$$I(q) \sim \int N(M)\, I_M(q)\, dM = \int N(M)\, M^2\, F(qR)\, dM \qquad (3.15).$$

Taking into account that a cluster with radius R has a mass $M \sim R^D$ and that the scattered momentum q due to this cluster is given by $q \sim 1/R$ we get $M \sim 1/q^D$. This is an important relation that will be used to perform the integral (3.15).

A particular situation where a large polydispersivity is present and has been intensively studied is the gelation process (sol-gel transition) described by the percolation models [18,35,36]. One of the wonderful things about gelation is that the incipient gel is a self-similar distribution of self-similar cluster, from monomers to the infinite cluster [37]. During the gelation process of branched polymers, there is a very large distribution number $N(M)$ of polymers with mass M. The distribution $N(M)$ obeys [38] an *exponential law* $P(M) = M^{-\tau} h(M/M_{av})$ where $M_{av}$ is the average mass of the polymers and $h(x)$ an exponentially decaying function of x which accounts for the cut-off of the distribution at large M. Near the percolation threshold $M_{av} \sim |p - p_c|^{-\gamma}$ where p is order parameter. The parameters $\tau$ and $\gamma$ are universal critical exponents given by the percolation theory [38].

In a first approximation let us neglect the cut-off function $h(x)$, putting $N(M) \sim M^{-\tau}$ and integrate (3.15) with M going from 0 up to $\infty$. In order assure that all mass contributions are taken into account the divide the integral into two parts: from 0 up to $1/q^{-D}$ and (B) from $1/q^{-D}$ up to $\infty$. Case (A) satisfy $0 \leq qR \leq 1$ so that $F(qR) = 1 - O(q^2)$ and likewise the contributions to (B) satisfy $\infty > qR \geq 1$, so that $F(qR) \approx (qR)^{-\gamma}$. Note that at $q = 1/R$ the mass is given by $M \sim R^D = q^{-D}$. In these conditions the A and B integrals $I(q)$ are given by [32,33]

$$I_A(q) \sim \int_0^{q^{-D}} M^{2-\tau}\, dM \sim q^{-D(3-\tau)}$$

if $\tau < 3$, and  (3.16)

$$I_B(q) \sim \int_{q^{-D}}^{\infty} M^{2-\tau}\, (qR)^{-D}\, dM \sim q^{-D(3-\tau)}.$$

if $\tau > 2$.

In this way, for $qR_m \ll 1$, where $R_m$ is the maximum value of R,

$$I(q) \sim q^{-D(3-\tau)} [1 - (qR_m)^2/3] \quad , \text{with } \tau < 3, \quad (3.17)$$

and for $qR_m \gg 1$

$$I(q) \sim q^{-D(3-\tau)} \quad , \text{with } \tau > 2 \quad (3.18).$$

Equations (3.16) and (3.17) show that for $qR_m \ll 1$ and $qR_m \gg 1$ in the log I(q) x q plot we have a straight line with a slope $(3-\tau)D$ instead of D.

Bouchaud et al.[33] have measured the scattered intensity for dilute polydispersive sample for large size polymers. They have found the slope $(3-\tau)D = 1.59$, in excellent agreement with the percolation theory that predicts $D = 2$ and $\tau = 2.2$. A review about the gelation process (sol-gel transition), where theories and experiments are contemplated, was done by Martin and Adolf [39]. In this paper the polydispersive case is also analyzed showing that predicted scattering intensity $I(q) \sim q^{-(3-\tau)D}$ is confirmed by many experimental results. According to usual percolation models[38] $\tau > 2$, however, models of kinetic aggregation [40] predict $\tau < 2$, becoming necessary to use a cut-off function $h(M/M_{av})$ to calculate (4.3.6). This case is discussed is details by Martin [32].

Martin and Ackerman [31,32] have also calculated I(q) taking into account the fractality of the clusters surfaces. They have shown that when the surface fractality is dominant for the limit $qR \gg 1$ we have

$$I_M(q) \sim M^2 (qR)^{-\gamma_s} \sim q^{-\gamma_s} \quad ,$$

where $\gamma_s = 6 - D_s$. If no cut-off function is used calculate (3.15) they have obtained

$$I(q) \sim q^{-3(3-\tau)} ,$$

showing that the scattering exponent for polydispersive system does not depend on the surface fractal dimension $D_s$ but only on the polydispersivity exponent $\tau$.

From the above results we verify that for the mass fractals I(q) depends on both parameters D and $\tau$, which seems reasonable enough. On the other hand, for surface fractals the power law polydispersivity destroys all information about the surface fractal dimension. However, when a cut-off function is introduced in the integral (3.15) it is shown [32] that the $\gamma_s$ factor of the exponential law depends on the $\tau$ values. Details of these predictions are shown by Martin [32].

**Acknowledgements.**
　　This work was supported by the Fundação de Amparo a Pesquisa do Estado de São Paulo (FAPESP) and the Conselho Nacional de Desenvolvimento Científico e Tecnológico (CNPq), Brazil. We also thank the librarian Virginia P. Franceschelli for the support at the IFUSP library.

**APPENDIX. Rayleigh calculation of $I_1(q)$.**

According to (2.1) the scattered intensity $I_1(q)$ due to an isolated monomer is given by

$$I_1(q) = I_e(q) \int_v d\mathbf{r} \int_v d\mathbf{r}'\, \rho(\mathbf{r})\, \rho(\mathbf{r}')\, \exp[i\mathbf{q}\cdot(\mathbf{r} - \mathbf{r}')] \quad (A.1),$$

where v is monomer volume and $\rho(\mathbf{r})$ the electronic density of the monomer. Indicating $<\rho(\mathbf{r})> = \rho$ the average electronic density we can write $\rho(\mathbf{r}) = \rho + \chi(\mathbf{r})$, where $\chi(\mathbf{r})$ represents the electronic density fluctuations of the monomer. In this way (A.1) becomes given by

$$I_1(q) = I_e(q) \int_v d\mathbf{r} \int_v d\mathbf{r}'\, [\rho + \chi(\mathbf{r})][\rho + \chi(\mathbf{r}')]\, \exp[i\mathbf{q}\cdot(\mathbf{r} - \mathbf{r}')] \quad (A.2),$$

where $I_e(q)$ is the intensity scattered by one electron and v the monomer volume.

Using (A.2) it can be shown [20,21] that

$$I(q) = I_e(q) \int_v d\mathbf{r} \int_v d\mathbf{r}'\, \chi(\mathbf{r})\, \chi(\mathbf{r}')\, \exp[i\mathbf{q}\cdot(\mathbf{r} - \mathbf{r}')] \quad (A.3).$$

Now, putting according to Guinier [12]

$$(1/v) \int_v \rho(\mathbf{r})\, \rho(\mathbf{r})\, d\mathbf{r} = (1/v) \int_v \rho(\mathbf{r})^2 d\mathbf{r} = <\rho^2> = \rho^2 \quad ,$$

where the brackets $<f>$ means an average of the variable f we define the *correlation function* $\gamma(\mathbf{r})$ by the relation

$$\int_v \chi(\mathbf{r})\, \chi(\mathbf{r} + \mathbf{x})\, d\mathbf{x} = \rho^2\, v\, \gamma(\mathbf{r}) \quad (A.4).$$

So, the intensity (A.3) becomes, omitting for simplicity the factor $I_e(q)$:

$$I_1(q) = \rho^2\, v \int_v d\mathbf{r}\, \gamma(\mathbf{r})\, \exp(i\mathbf{q}\cdot\mathbf{r}) \quad (A.5).$$

For spherical particles with radius $r_o$ it can be shown [21,12,13] that the function $\gamma(r)$ shown in (2.3) is given by

$$\gamma(r) = \gamma_o(r) = 1 - 3r/4r_o + (1/16)(r/r_o)^3 \qquad (A.6).$$

Substituting (A.6) into (A.5) one can calculate the intensity $I_1(q)$. It is shown [12,13] that the $I_1(q)$ predictions obtained in this way are quite similar to that found using (2.3).